# Photodetecting and Light-Emitting Devices Based on Two Dimensional Materials

Yuanfang Yu[1], Feng Miao[2,*], Jun He[3,*], Zhenhua Ni[1,*]

[1]*Department of Physics and Key Laboratory of MEMS of the Ministry of Education, Southeast University, Nanjing 211189, China.*
[2]*National Laboratory of Solid State Microstructures, School of Physics, Collaborative Innovation Center of Advanced Microstructures, Nanjing University, Nanjing 210093, China.*
[3]*CAS Key Laboratory of Nanosystem and Hierarchical Fabrication, National Center for Nanoscience and Technology, Beijing 100190, China.*

**Abstract:**

Two dimensional (2D) materials, e.g. graphene, transition metal dichalcogenides (TMDs), black phosphorus (BP), have demonstrated fascinating electrical and optical characteristics and exhibited great potential in optoelectronic applications. High performance and multifunctional devices were achieved by employing diverse designs of architectures, such as hybrid systems with nanostructured materials, bulk semiconductors and organics, forming 2D heterostructures. In this review, we mainly discuss the recent progresses of 2D materials in high responsive photodetectors, light-emitting devices and single photon emitters. Hybrid systems and van der Waals heterostructures based devices are emphasized, which exhibit great potential in state-of-the-art applications.

Corresponding Authors: Zhenhua Ni (zhni@seu.edu.cn); Feng Miao (miao@nju.edu.cn); Jun He (hej@nanoctr.cn).

# 1. Introduction

2D materials possess superior optical, electrical, mechanical properties and have attracted great attentions in the past decade.[1-5] Graphene, as a precedent discovered character in 2D materials, has ultrahigh carrier mobility and broadband light absorption.[1,6] Charge carrier density in graphene can be effectively tuned, which enables graphene a promising candidate for electrically tunable devices. Different from the gapless graphene, 2D TMDs, such as $MoS_2$ and $WSe_2$, are semiconductors with bandgap ranging from 1.57 to 2 eV,[7] which are more suitable for applications such as photodetectors and light emitting diodes (LEDs). BP, a new member in 2D materials family, exhibits direct bandgap of about ~1.8 and ~0.3 eV in its single layer and bulk form, respectively.[8-11] The bandgap of 2D semiconductors can be effectively tuned by their thickness and the properties will change accordingly, which meet the demands of various applications. Besides the 2D layered materials mentioned above, 2D non-layered materials, such as PbS and $Pb_{1-x}Sn_xSe$, also exhibit great potential in the application of optoelectronic devices. PbS has direct narrow bandgap of ~0.4 eV, which is available for broad spectral detection from visible to mid-infrared region.[12,13] $Pb_{1-x}Sn_xSe$ is also a direct narrow band gap semiconductor with great potential in infrared detection and thermoelectric conversion.[14,15] The strong light-matter interaction, broad range light absorption/emission, and high carrier mobility of 2D atomically-thin materials make them promising candidates in optoelectronic applications, such as photodetectors and light-emitting devices.

It is crucial to exploit photodetectors with high performance in terms of speed, wavelength range and sensitivity. However, high responsive photodetection is a challenge for graphene, due to its relatively weak light absorption (~2.3% per layer).[16] This problem can be solved effectively by employing hybrid architecture and the photogating mechanism, which can remarkably improve the photosensitivity by combining graphene with other nanomaterials, such as quantum dots (QDs). On the other hand, PN junction is a key construction in optoelectronic applications, such as photodetectors and LEDs. The junctions were commonly formed by chemical doping[17] and electrostatic gating[18,19]. Besides the traditional approaches, van der Waals heterostructures based on 2D materials are ideal choices for achieving state-of-the-art devices, which provides a platform for combining superiorities of different 2D layered materials, and also forming different types of junctions. In addition, single photon emitters are another important part of applications for 2D materials, which bring new opportunities to quantum technologies.

In this review, we will discuss the recent progress on photodetecting and light-emitting applications of 2D materials. In section 2, we mainly review a series of state-of-the-art photodetectors based on 2D materials, including graphene, TMDs, BP and 2D non-layered materials. Devices based on hybrid systems and van der Waals heterostructures are emphasized. In section 3, the light-emitting applications of 2D materials are summarized and electroluminescence in well-designed LEDs is presented. Subsequently, single photon emission originating from the defects in 2D materials is discussed. Finally, we discuss the perspective and challenges of 2D materials based optoelectronics devices.

## 2. Photodetectors based on 2D materials

### 2.1 Mechanisms of photodetection in 2D materials

Photodetectors convert optical signals to electrical signals, which are broadly applied in communications, imaging systems and remote sensing.[20-23] Here, we simply discuss the photovoltaic and photogating effects, which are the most commonly adopted operating mechanisms for high speed and high responsive photodetection, respectively.

In photovoltaic detectors, the photogenerated electron-hole pairs in the channel of detectors are separated with the help of the built-in electric field at the junctions (e.g. PN junctions, Schottky junctions), thus produce photocurrent. Junctions can be formed by different approaches, for instance, by chemical doping,[17] electrostatic gating,[18,19] or combining two materials which possess different work functions[24]. Due to the ultrafast generation and separation of electron-hole pairs at the junctions, the photodetectors dominated by photovoltaic effect are capable for ultrafast photodetection.[25] In photogating devices, take graphene based photodetectors for example, the incident light produces electron-hole pairs in photon absorbers (e.g. QDs), with one of the two types of carriers transferred to graphene and another one trapped and gate the graphene channel. Driving by bias voltage, the type of carrier transferred to graphene would recirculate between source and drain, and produce ultrahigh gain and photoresponsivity. Besides, there are other mechanisms in photodetectors, such as photo-thermoelectric effect[26,27] and bolometric effect[28], which are not discussed here.

### 2.2 Hybrid photodetectors based on graphene

The ultra-broadband light absorption from ultraviolet to terahertz (THz) enables graphene a promising candidate for broadband photodetection. However, the responsivity of a graphene based photodetector is limited by the relatively weak light absorption and ultrashort lifetime of photon-induced carriers. In order to meet the demands of high responsive devices, hybrid structures with photon absorbing materials are frequently employed.[29-31] For instance, combining QDs and graphene can effectively prolongs the life-time of photon-induced carriers and thus realizes high responsive photodetection.[32] Hybrid phototransistors based on PbS QDs and graphene were demonstrated with ultrahigh gain and a high responsivity of ~$10^7$ A $W^{-1}$.[32] By employing lead zirconatetitanate ($Pb(Zr_{0.2}Ti_{0.8})O_3$) (PZT) substrates instead of silicon dioxide ($SiO_2$), responsivity as high as $4.06\times10^9$ A $W^{-1}$ under 325 nm light illumination was achieved in a composite graphene and graphene quantum dots (GQDs) photodetector.[33] Besides QDs and graphene hybrid photodetectors, a series of novel hybrid structures have also been investigated, which will be discussed in detail below.

### 2.2.1 Hybrid systems with bulk semiconductors

The combination of silicon and graphene is a typical example for bulk semiconductors/graphene hybrid structure.[24,34,35] In Fig. 1(a), lightly p-doped silicon was employed to form a silicon/graphene heterostructure.[34] Responsivity of such photodetector can exceed $10^4$ A $W^{-1}$ at wavelength of 632 nm (shown in bottom left graph of Fig. 1(a)) and reach 0.23 A $W^{-1}$ at 1550 nm (shown in bottom right graph of Fig. 1(a)). In the visible region, there are three sources of photo-induced carriers: the interband transition in graphene, the depletion region in the Schottky junction (at silicon/graphene interfaces), and the bulk region of silicon. The absorption of light by silicon can result in injection of carriers into graphene, and the large recombination time scale and the ultrafast transition of carriers within graphene lead to ultrahigh gain of the injected carriers and thereby the high responsivity of the devices. In the infrared region, graphene is the only light absorber, and the relatively smaller absorption and shorter recombination time lead to low responsivity. In another silicon/graphene hybrid photodetector (Fig. 1(b)), ultrahigh quantum gain exceeds $10^6$ electrons per incident photon and responsivity of nearly $10^7$ A $W^{-1}$ were realized.[35] The quantum carrier reinvestment was proposed to explain the detail mechanism. Excellent weak-signal detection based on silicon/graphene heterojunctions with photovoltage responsivity exceeding $10^7$ V $W^{-1}$ was also reported.[24] The manipulation of Fermi level of graphene enables a high degree of tunability and efficient capture of photoexcited carriers. There are also other bulk semiconductors employed in graphene based hybrid photodetectors, e.g. germanium.[36]

Besides forming silicon/graphene junctions, interfacial gating effect was also utilized to achieve high performance photodetection. An ultra-fast and high sensitive photodetector was achieved by employing graphene/SiO$_2$/lightly-doped-Si hybrid architecture with interfacial gating mechanism (Fig.1(c)).[37] The localized interface states at the oxide-silicon interface induce a negative depletion layer in the silicon near the interface and cause surface energy bands to bend downwards, leading to the formation of thermal equilibrium and built-in electric field.[38] Photogenerated electron-hole pairs in silicon are separated, the holes diffusing to silicon bulk and the electrons remaining at the SiO$_2$/silicon interface, forming a negative voltage on the interface and thereby gating the graphene channel through capacitive coupling. The holes carrier density of graphene increased due to gating effect, and thus high photocurrent was achieved. The detector exhibits superior capability on weak signal detection, with high responsivity of ~1000 A W$^{-1}$ for weak signal of <1 nW (shown in the right graph of Fig. 1(c)). Most importantly, the response time of the device has been pushed to ~400 ns, due to the ultra-fast separation of electron-hole pairs at the interface. Heavily doped silicon was not available due to its very short lifetime of the photogenerated electron-hole pairs, for which the negative voltage in the interface can be approximately neglect.[39] The interfacial gating effect opens a new route for high responsivity in semiconductor/graphene hybrid devices.

### 2.2.2 Hybrid systems with organic crystals and perovskite

Except for bulk semiconductors, other light absorbing materials have also been employed in graphene based hybrid photodetectors, such as organic crystals and organic-inorganic hybrid perovskite.[40-44] An ultrathin epitaxial organic crystal/graphene hybrid structure phototransistor is presented in Fig. 2(a).[40] Photogenerated electron-hole pairs in C$_8$-BTBT are separated by built-in electric field formed at the interface of graphene and C$_8$-BTBT, and electrons move toward graphene. Photoresponse of the device was studied with increasing C$_8$-BTBT thickness. In few-layer C$_8$-BTBT devices, the responsivity could reach 4.76×10$^5$ A W$^{-1}$.[40] CH$_3$NH$_3$P$_b$I$_3$ perovskite/graphene hybrid system was also employed for photodetection.[41] Electrons in graphene transferred to the adjacent perovskite layer and filled the empty states in the perovskite valence band. As a result, the photoexcited electrons in the perovskite reside in the conduction band without decaying. The trapped electrons produced effective photogating effects, so that the presence of the charges altered the conductivity of the graphene channel through capacitive coupling.[45,46] The photodetector exhibits a photoresponsivity of 180 A W$^{-1}$ and an external quantum efficiency (EQE) of ~5×10$^4$% for an optical power on the microwatts scale (Fig. 2(b)), with photodetectivity of ~10$^9$ Jones. There are other organic materials employed in organics/graphene hybrid photodetectors and achieved high performance, e.g. poly(3-hexylthiophene-2,5-diyl) (P3HT)[42,43] and tetraphenyl-porphyrin (H$_2$TPP).[44]

### 2.2.3 Hybrid devices for infrared photodetection

Infrared photodetection plays an important role in the field of communications, imaging and astronomy.[20,21,47] 2D materials are easy to be integrated in electronic devices, and many of them exhibit considerable photoresponse in infrared range. Hybrid architecture has been widely utilized in 2D materials based infrared photodetectors.

Carbon nanotubes (CNTs) exhibit great potential in infrared photodetection, due to their high absorption coefficient in the infrared spectrum.[48] High detectivity up to 1.5×10$^7$ cm Hz$^{1/2}$W$^{-1}$ has been demonstrated in a multiwall carbon nanotube (MWCNT)/graphene hybrid photodetector, which presents a 500% improvement over the best photodetectivity achieved on MWCNT film infrared detectors.[49] Single-wall carbon nanotubes (SWCNT)/graphene hybrid film was also employed to form a large built-in potential at the interface, which performs well in the separation of electron-hole pairs and decreases the recombination of spatially isolated photocarriers.[50] The hybrid photodetector (shown as inset in the top panel of Fig. 3(a)) operates in photogating mechanism and demonstrates a photoconductive gain of 10$^5$ with ultra-broadband sensitivity from visible to near-infrared (400-1550 nm). It exhibits great performance with a high photoresponsivity of over 100 A W$^{-1}$ and a fast response time of 100 μs.

Topological insulators, e.g. Bi$_2$Te$_3$, Bi$_2$Se$_3$ and Sb$_2$Te$_3$, which generally have a very small band gap (0.15-0.3 eV) in the mid-infrared range, are good candidates for infrared photodetection.[51,52] A photodetector based on Bi$_2$Te$_3$/graphene heterostructure was demonstrated (Fig. 3(b))[31], which enables valid transfer and separation of photo-excited carriers at the Bi$_2$Te$_3$/graphene interface. Photocurrent of the device can be effectively enhanced without sacrificing the detecting spectral width. The device exhibited high responsivity from visible to infrared range (400-1550 nm).(Fig. 3(b) bottom)

Different from the hybrid photodetectors mentioned above, a broadband photodetector based on graphene double-layer heterostructure was presented, in which the photo absorbing and electron conductive layers are both using graphene.[53] Hot electrons and holes are separated into opposite graphene layers by selective quantum tunneling, thus hot carrier recombination was minimized. The trapped charges on the top graphene layer can lead to a strong photogating effect on the bottom graphene channel layer, yielding a considerable photoresponsivity over an ultra-broad spectral range (from visible to mid-infrared range). The photoresponsivity is as high as 1.1 A $W^{-1}$ at the wavelength of 3.2 um. The hot carrier tunneling mechanism in the graphene double-layer heterostructure provides an available route for ultra-broadband and high-sensitivity photodetection at room temperature.

### 2.3 Photodetectors based on TMDs and BP

### 2.3.1 Photodetectors based on TMDs

2D TMDs are well known for their remarkable light absorption properties, with energy gaps in the near-infrared to the visible spectral region. High performance photodetectors based on different types of TMDs were successfully demonstrated. Single-layer $MoS_2$ is a direct-gap semiconductor with a bandgap of ~1.8 eV.[54] The direct bandgap would enable efficient light absorption and electron-hole pair generation under photoexcitation. In a previous work, $MoS_2$ was used as the channel material in a field-effect transistor and exhibited a high channel mobility (~200 $cm^2V^{-1}S^{-1}$) and current ON/OFF ratio ($10^8$).[55] With improved mobility, as well as the contact quality and positioning technique, high responsive monolayer $MoS_2$ based photodetector was realized with a maximum photoresponsivity of ~880 A $W^{-1}$ at a wavelength of 561 nm (Fig.4(a)).[56] Due to the short electron-hole and exciton lifetimes in $MoS_2$, intrinsic response times as short as 3 ps was realized in a monolayer $MoS_2$ photodetector (bandwidths of ~300 GHz) (Fig. 4(b)).[57] The photo response time characterization (Fig. 4(b) right) actually shows two distinct timescales: (i) a fast timescale of ~4.3 ps and (ii) a slow timescale of ~105 ps. Such phenomenon is explained by a model considering carrier capture by two different defect levels (fast and slow defects). After photoexcitation, electrons and holes in $MoS_2$ thermalize and lose most of their energy within a short time. Most of the photoexcited holes and then most of the electrons, are captured by the fast defects within the first few ps. Meanwhile, the minority of the holes is also captured by the slow defects.

$ReS_2$ is another member of TMDs, which has a direct bandgap of ~1.5 eV and exhibits anisotropic nature within the layer plane.[58] Both single and muiltlayer $ReS_2$ are direct bandgap due to the weak interlayer coupling.[59] Few layer $ReS_2$ has a higher density of states[60] and stronger light absorption than the thinner one.[54] Ultrahigh responsivity (~88600 A $W^{-1}$) phototransistors based on few layer $ReS_2$ were demonstrated (shown in Fig. 4(c) and Fig. 4(d)), which is promising for weak signal detection.[61] The impurities or S vacancies in ReS$_2$ flakes act as trap states, the electrons of the photogenerated electron-hole pairs stay at the trap states for a long time and thereby present a long lifetime of holes, thus lead to the increase of the photoconductive gain. Furthermore, the use of thicker ReS$_2$ could increase the absorbed optical power and hence lead to large quantum efficiency.

The dark current would strongly affect the performance of photodetectors, e.g. on/off ratio. Novel device architectures have been adopted in TMDs based photodetectors to suppress the dark current and improve the device performance. Ferroelectric materials combined with 2D materials have been proposed for photodetectors, with a poly(vinylidene fluoride-trifluoroethylene) (P(VDF-TrFE)) ferroelectric polymer film covered on a MoS$_2$ effect transistor (Fig. 5(a)).[62] The ferroelectric polarization could strongly suppress the darkcurrent of the photodetector. The depleted state of carriers in the MoS$_2$ channel is caused by the electrostatic field derived from the remnant polarization of P(VDF-TrFE). In polarization up state, the drain-source current is the lowest compared to those of the other two states: without polarization and polarization down state. Furthermore, the band structure of few-layer MoS$_2$ can also be modified by the effect of the strong electric field polarization. The photoresponse wavelengths of the device were then extended from the visible to the near-infrared (0.85-1.55 μm). Such ferroelectric/photoelectric 2D material hybrid system is a promising strategy for high performance 2D electronic/optoelectronic devices. In another work, gold nanoparticles (AuNPs) were employed in a WS$_2$ phototransistor, as shown in Fig. 5(b).[63] Efficient electron trapping originated from AuNPs embedded in the gate dielectric could strongly suppress dark current. Such a device exhibited ultralow dark current ($10^{-11}$A), high photoresponsivity (1090 A W$^{-1}$) and high detectivity (3.5 × $10^{11}$ Jones) at a wavelength of 520 nm with a low bias voltage and a zero gate voltage.

**2.3.2 Photodetectors based on BP**

As a new member in 2D materials family, BP is an anisotropic material, which exhibits a high Hall mobility up to 1,000 cm$^2$V$^{-1}$s$^{-1}$ at room temperature.[64] BP exhibits direct bandgap of about ~1.8 and ~0.3eV in its single layer and bulk form, respectively,[8-11] and tunable bandgap with different thicknesses covering the visible to mid-infrared spectral range. High responsivity of about $10^3$ A W$^{-1}$ at 300 K and 7×$10^6$ A W$^{-1}$ at 20 K in the near-infrared region (900 nm) has been achieved in a BP based photodetector, as shown in Fig. 6(a).[65] The contact metal Ni is employed and forms a good ohmic contact to BP under p-type operation, results in efficient collection of the photogenerated carriers. The small contact resistance of the device greatly improves the on current and thus the photocurrent. Due to the strong intrinsic linear dichroism, BP photodetector exhibits great polarization sensitivity over a broad wavelength from 400 to 3750 nm (Fig. 6(b)).[66] To further improve the performance, the BP flakes were patterned into an ionic gel gated electric-double-layer transistor (EDLT), which is able to tune interfacial band bending and also the Fermi level of channel materials over a large range.[67] Photogenerated electrons and holes are separated by selectively driving them into surface or bulk layers under a built-in electric field within an EDLT. Therefore the efficiency of linear dichroism photodetector can be greatly enhanced.[66] BP was also integrated with silicon waveguide to realize high responsivity photodetection,[68] as shown in Fig. 6(c). The photocurrent was dominated by photovoltaic current at low doping case and the bolometric effect at high n-type doping case. Intrinsic responsivity reaches 135 mA $W^{-1}$ (thickness of BP 11.5 nm) and 657 mA $W^{-1}$ (thickness of BP 100 nm) at room temperature, with a high response bandwidth exceeding 3 GHz (as shown in the right panel of Fig. 6(c)).

### 2.4 Photodetectors based on 2D heterostructure

Van der Waals heterostructure formed by 2D materials is commonly adopted for high performance photodetection, since it provides a platform for combining superiorities of different 2D layered materials.

Coupling graphene with $MoS_2$ can produce a hybrid material that combine the high photon absorption capability of $MoS_2$ and high carrier mobility of graphene. A photodetector based on graphene/$MoS_2$ heterostructure exhibits a high photogain greater than $10^8$ and a photoresponsivity value higher than $10^7$ A $W^{-1}$ (as shown in Fig. 7(a)).[69] In graphene/$MoS_2$ heterostructure, the electron-hole pairs are produced in the $MoS_2$ layer upon light illumination, and separated at the $MoS_2$ and graphene interfaces. The electrons can move to the graphene layer due to an effective built-in electric field, while the holes are trapped in the $MoS_2$ layer. The high electron mobility in graphene and the long charge-trapping lifetime of the holes leading to multiple recirculation of electrons in graphene, result in a very high photogain. Photodetector based on graphene-$MoS_2$-graphene vertical heterostructure was also demonstrated, as shown in Fig.7(b).[70] The amplitude and polarity of the photocurrent in the gated vertical heterostructures can be easily modulated by the electric field of an external gate, and the maximum EQE of 55% and internal quantum efficiency up to 85% was achieved. Heterostructures based on different types of TMDs are also widely investigated. The gate-tunable diode-like current rectification and a photovoltaic response across the p-n interface have been observed in p-type $WSe_2$ and n-type $MoS_2$ atomically thin vertical p-n junction, as shown in Fig. 7(c).[71] The photocurrent effectively tuned by gate voltages was shown in the bottom panel in Fig. 7(c). It was found that the interlayer tunneling recombination of the majority carriers across the van der Waals interface, which can be tuned by gating, can significantly influence both the electrical and optoelectronic properties of the junction. There are also more complicated heterostructures, e.g. a photodetector based on $MoS_2$-graphene-$WSe_2$ heterostructure, as shown in Fig. 7(d).[72] Photoexcited electrons and holes were efficiently separated by the built-in electric field formed at the depletion region of the p-n junction, and enables broadband photodetection (from the visible to infrared range) with high sensitivity. In the visible range, the photon energy is larger than the bandgaps of the TMDs materials, and abundant photogenerated free carriers are produced by all the three layered materials, resulting in a considerably higher photoresponse. When the wavelength is near the infrared region and the photon energy is smaller than the bandgap of TMDs, the interband absorption of both monolayer $MoS_2$ and $WSe_2$ is forbidden. In such case, graphene is the only photon absorbing medium, leading to a relatively smaller photoresponse of the heterostructure in the infrared range. The device exhibits outstanding performance with photoresponsivity of $10^4$ A $W^{-1}$ at 400 nm, and the specific detectivity reaches $10^{15}$ Jones and $10^{11}$ Jones in the visible and near-infrared region, respectively.

### 2.5 Photodetectors based on 2D non-layered materials

2D layered materials have strong lateral chemical bonding in planes but weak van der Waals interaction between planes. On the other hand, many other materials with non-layered nature of their bulk crystals can be produced as ultrathin nanosheets and nanoplates.[73] These 2D non-layered materials, such as SnTe, $Pb_{1-x}Sn_xSe$, and $Pn_{1-x}Sn_xTe$, possess narrow bandgap and are promising candidate for infrared photodetection.[ 14,74-76]

$Pb_{1-x}Sn_xSe$ nanoplates with thickness 15 to 45 nm can be fabricated from their bulk crystals, as illustrated in Fig. 8(a).[77] The crystal morphology of $Pb_{1-x}Sn_xSe$ is shown in Fig. 8(b). Flexible photodetectors based on $Pb_{1-x}Sn_xSe$ nanoplates exhibit fast, reversible, and stable photoresponse, and broad spectra detection ranging from UV to infrared. High-performance mid-infrared detector based on $Pb_{1-x}Sn_xSe$ nanoplates with detection wavelength extending to 1.7-2.0 μm is shown in Fig.

8(c).[78] The photoresponsivity is estimated to be 318 mA W$^{-1}$ at the wavelength of 1.9-2.0 μm, shows great potential for applications in military communication, remote sensing and environmental monitoring. PbS is a semiconductor with direct electronic bandgap of ~0.4eV.[12] 2D PbS nanoplates can be successfully grown by chemical vapor deposition (CVD) method, and the thickness ranges from 5 to 35 nm.[13] The devices based on 2D PbS nanoplates exhibit remarkable infrared response with high photoresponsivity (1621 AW$^{-1}$), detectivity (1.72×10$^{11}$ Jones) and photogain (2512) in the wavelength of 800 nm (Fig. 8(d)).[13] 2D ultrathin metal oxide nanosheets, including $TiO_2$, ZnO, $Co_3O_4$ and $WO_3$, are also fabricated and utilized in photodetection.[79] The metal oxide nanosheets based devices exhibit considerable photoresponse and present excellent stability. The above results do demonstrate that photodetectors based on 2D non-layered materials exhibit excellent performance in broadband photodetection, especially in the infrared range.

## 3. Light emitting applications of 2D materials

Single-layer TMDs emerged as promising candidates for light emitting devices, which possess direct bandgap electronic band structure and high light emitting efficiency. The light emission properties are dominated by exciton effect in 2D TMDs.[80-82] Due to the strong Coulomb force interaction, charged carriers generated optically or electrically in TMDs form different kinds of combination models, such as exciton (an electron-hole pair), trion (two electrons with a hole, or two holes with an electron), bi-exciton (two electrons with two holes).[83] In these models, exciton and bi-exciton exhibit charge-neutral, whereas trion expresses charged feature. Free exciton can also be bound within potential well, hence form bound exciton.[84] Electroluminescence is a physical phenomenon that converts electrical energy to optical energy, which emits lights originating from the electron-hole recombination. In TMDs, p or n types doping can be easily realized, which benefit in constructing junctions for LEDs. Following, we will discuss LEDs based on TMDs and heterostructures.

### 3.1 LEDs based on TMDs

Employing two local gates to define a p-n junction within the TMDs sheet is the most commonly used way to build LEDs.[85,86] The LEDs based on p-n junction in $WSe_2$ show outstanding performance with total photon emission rate up to ~16 million s$^{-1}$ at applied current of 35 nA (Fig. 9(a)).[87] In the right panel of Fig. 9(a), from left to right, the arrows indicate electroluminescence features that correspond to the impurity-bound exciton, two types of the charged excitons and the neutral exciton, respectively. Ionic liquid gated structure is known with high gating level, and can tune the Fermi level of channel materials over a large range. Electroluminescence has also been observed in atomically thin ionic-liquid gated light-emitting transistors, as shown in Fig. 9(b).[88] Effective electron and hole accumulation in the device enables the operation of the transistors in the ambipolar injection regime, with electrons and holes injected simultaneously at the two opposite contacts of the devices, and lead to light emission. The electro- and the photo- luminescence spectra are similar to each other in both mono- and the bilayer devices (shown in the bottom right panel of Fig. 9(b)). Polarized $WSe_2$-based LEDs was also reported, which can emit circularly polarized electroluminescence from p-i-n junction, as shown in Fig. 9(c).[89] The circularly polarized electroluminescence is attributed to the generation of valley-polarized charge currents due to trigonal warping under high electrical bias and the valley-dependent optical selection rules. The degree of circular polarization reaches 45%, and remarkably, the polarization can be electrically controlled.

### 3.2 LEDs based on heterostructure

Heterostructures based on TMDs have also been developed to for LEDs. A diode of monolayer $MoS_2$ fabricated on a heavily p-type doped silicon substrate is shown in Fig. 10(a).[90] When a forward bias is applied to the $MoS_2$/silicon heterojunction, due to the direct band-gap of monolayer $MoS_2$, the injection of holes from silicon across the junction can give rise to efficient radiative recombination. The electroluminescence spectra at room temperature and low temperature are displayed in (Fig. 10(a) bottom), two Lorentz contributions fitted in the spectra are attributed to the A exciton (labeled AX) and bound exciton (labeled DX) emission. The high signal-to-noise ratio makes it possible to identify the emission from different optical transitions. LED made with a p−i−n heterojunction of p+-Si/CVD-grown i-$WS_2$/n-ITO has also been reported, as shown in Fig. 10(b).[91] The heavily p-doped Si and n-ITO are adopted to inject holes and electrons into i-$WS_2$, respectively. The circular polarization degree of the device reaches up to 81% and can be modulated by forward current. Vertical van der Waals heterostructures have also been employed in LEDs, which enables effective charge transfer across the atomically sharp heterojunctions. $WSe_2$/$MoS_2$ heterojunction p−n diode is presented in Fig. 10(c), where the p−n junction is created over the entire $WSe_2$/$MoS_2$ overlapping area.[92] $WSe_2$/$MoS_2$ heterojunction based LEDs show prominent band edge excitonic emission and strikingly enhanced hot-electron luminescence. Such a novel heterostructure system opens up a new route to novel optoelectronic devices including atomically thin photodetectors as well as spin- or valley-polarized LEDs. Quantum wells (QW) can be formed by vertical heterostructures structure, as shown in Fig. 10(d).[93] Such a vertical heterostructures based LEDs is superior to lateral p-n junctions based devices in many aspects, including reduced contact resistance, higher current densities as well as the large luminescence area. Electrons and holes are injected into TMDs from two graphene electrodes, then recombine and emit a photon, due to the long lifetime of the quasiparticles in the QWs. The LEDs based on single quantum well present quantum efficiency above 1% and line widths of 18 meV. Furthermore, by utilizing multiple quantum wells, EQE of the device can reach 8.4%.

**3.3 Single photon emitters based on 2D materials**

A high-quality and high-efficiency single-photon source is required for implementing photonic quantum information processing and quantum key distribution.[94] Crystal structure imperfections can act as sources of single photon emission when they are isolated appropriately. Recently, single-photon emission from 2D materials, including $WSe_2$ and hexagonal boron nitride (hBN), has been reported.[95-99]

Several groups[95-98] have reported single-photon emitters in 2D $WSe_2$, with one of the examples shown in Fig. 11(a).[95] The emitters based on excitons spatially localized by defects in monolayer $WSe_2$ show very narrow linewidths of ~130 μeV. Through second-order correlation measurements, the strong photon antibunching is exhibited and the single-photon nature of the emission is confirmed (shown in the right panel of Fig. 11(a)). Effectively controllable single photon emission will be beneficial for quantum information processing. The single photon emission properties in $WSe_2$ can be controlled by external direct current electric and magnetic fields manipulation (Fig. 11(b)).[97] For example, Zeeman splitting is exhibited in many of these single photon emitters as displayed in the right panel of Fig. 11(b). Single-photon emission was also demonstrated in hexagonal boron nitride (BN) monolayer and multilayers at room temperature,[99] which originate from a defect that a boron atom is replaced by a nitrogen atom adjacent to vacancy. The emitters showed maximum emission rate of $I_\infty \approx 4.2 \times 10^6$ counts s$^{-1}$ at the saturated excitation power about 611 μW, as illustrated in Fig. 11(c).

The single photon emission demonstrated in 2D material is beneficial to the development quantum technology, especially for the emitters operating at room temperature[99]. Compared with the traditional solid-state single photon emitters, which are typically embedded in bulk materials, 2D materials are easier to be integrated into electronic devices and their emission properties are easier to be controlled.

## 4. Outlook and perspective

In summary, great achievements have been realized in the rapidly developing field of 2D materials optoelectronics in recent years. High responsive graphene based photodetectors have been realized by employing different hybrid systems, including QDs, bulk semiconductors, organics, topological insulators, and so on. TMDs based photodetectors have also exhibit high performance in terms of high sensivity and high speed, especially in the visible region. BP, as the new member of 2D materials, was extensively investigated in broadband photodetection with polarization sensitivity. Besides the 2D layered materials, 2D non-layered materials, such as SnTe, $Pb_{1-x}Sn_xSe$, and $Pn_{1-x}Sn_xTe$, possess narrow bandgap and are promising candidate for infrared photodetection. Van der Waals heterostructures are able to combine the advantages of various 2D materials. By virtue of these constructions, strong light absorption and broadband response have been achieved. Many state-of-the-art optoelectronics, e.g. photodetectors and LEDs, are based on heterostructures. Single photon emitters have also been demonstrated in $WSe_2$ and hBN, with defects play very important role inside. However, more efforts are expected on but not restricted to the following aspects:

1). The responsivity, dark current level, response time are important parameters of photodetectors. High responsivity has been achieved in the photogating effect dominated photodetectors, but the price for achieving ultra-high sensitivity is sacrificing the response time. It is challenging to balance these parameters for specific applications.

2). Infrared photodetection plays an important role in communications, astronomy and military systems. Besides the broadly studied layered materials, e.g. graphene and BP, the 2D non-layered materials, such as SnTe, $Pb_{1-x}Sn_xSe$ and $Pn_{1-x}Sn_xTe$, possess narrow bandgap and are promising candidates for infrared photodetection. We expected that high quality 2D non-layered materials would be fabricated with well-designed techniques and high performance infrared photodetectors can further be achieved. Furthermore, hybrid system can be employed in 2D non-layered materials to enhance the light absorption and responsivity.

3).Considerable EQE (nearly 10%) has been achieved in vertical heterostructures based LEDs, which inspired us to explore more outstanding devices by band structure engineering in van der Waals heterostructure. It is also crucial to improve the electroluminescence efficiency by means of exploiting novel constructions and suitable materials, as well as modulating the properties of 2D materials.

4). Defects can strongly affect the properties of 2D materials. These imperfects, such as vacancies, can act as potential resources for light emission. It demands constant efforts to achieve high efficient LEDs and explore novel properties of different type of defects through defects engineering. It is expected that more single photon emissions originated from the defects in layered TMDs materials would be discovered in the future.

**Acknowledgements**

This work is supported by NSFC (61422503 and 61376104), the open research funds of Key Laboratory of MEMS of Ministry of Education (SEU, China), and the Fundamental Research Funds for the Central Universities.

**References**
[1] Mak K F, Ju L, Wang F and Heinz T F 2012 Solid State Commun. 152 1341
[2] Splendiani A, Sun L, Zhang Y B, Li T S, Kim J, Chim C Y, Galli G and Wang F 2010 Nano Lett. 10 1271
[3] Xu X D, Yao W, Xiao D and Heinz T F 2014 Nature Phys. 10 343
[4] Castro Neto A H, Guinea F, Peres N M R, Novoselov K S and Geim A K 2009 Rev. Mod. Phys. 81 109
[5] Geim A K 2009 Science 324 1530
[6] Novoselov K S, Geim A K, Morozov S V, Jiang D, Zhang Y, Dubonos S V, Grigorieva I V, Firsov A A 2004 Science 306 666
[7] Sun Z P, Martinez A and Wang F 2016 Nat. Photonics 10 227
[8] Du Y L, Ouyang C Y, Shi S Q and Lei M S 2010 J. Appl. Phys. 107 093718
[9] Akahama Y, Endo S and Narita S 1983 J. Phys. Soc. Jpn. 52 2148
[10] Das S, Zhang W, Demarteau M, Hoffmann A, Dubey M and Roelofs A 2014 Nano Lett. 14 5733
[11] Takao Y and Morita A 1981 Physica B+C 105 93
[12] Nichols P L, Liu Z, Yin L, Turkdogan S, Fan F and Ning C Z 2015 Nano lett. 15 909
[13] Wen Y, Wang Q S, Yin L, Liu Q, Wang F, Wang F M, Wang Z X, Liu K L, Xu K,

Huang Y, Shifa T A, Jiang C, Xiong J and He J 2016 Adv. Mater. DOI: 10.1002/adma.201602481

[14] Dziawa P, Kowalski B J, Dybko K, Buczko R, Szczerbakow A, Szot M, Łusakowska E, Balasubramanian T, Wojek B M, Berntsen M H, Tjernberg O and Story T 2012 Nat. Mater. 11 1023

[15] Dixon J R, Hoff G F 1971 Phys. Rev. B 3 4299

[16] Nair R R, Blake P, Grigorenko A N, Novoselov K S, Booth T J, Stauber T, Peres N M R and Geim A K 2008 Science 320 1308

[17] Farmer D B, Golizadeh-Mojarad R, Perebeinos V, Lin Y M, Tulevski G S, Tsang J C and Avouris P 2008 Nano Lett. 9 388

[18] Peters E C, Lee E J H, Burghard M and Kern K 2010 Appl. Phys. Lett. 97 193102

[19] Lemme M C, Koppens F H L, Falk A L, Rudner M S, Park H, Levitov L S, and Marcus C M 2011 Nano Lett. 11 4134

[20] Mueller T, Xia F N and Avouris P 2010 Nat. Photonics 4 297

[21] Rogalski A, Antoszewski J and Faraone L 2009 J. Appl. Phys. 105 091101

[22] Engel M, Steiner M and Avouris P 2014 Nano Lett. 14 6414

[23] Chen C J, Choi K K, Chang W H and Tsui D C 1998 Appl. Phys. Lett. 72 7

[24] An X H, Liu F Z, Jung Y J and Kar S 2013 Nano Lett. 13 909

[25] Tielrooij K J, Piatkowski L, Massicotte M, Woessner A, Ma Q, Lee Y, Myhro K S, Lau C N, Jarillo-Herrero P, vanHulst N F and Koppens F H L 2015 Nat. Nanotechnol. 10 437

[26] Wang W H, Nan H Y, Liu Q, Liang Z, Yu Z H, Liu F Y, Hu W D, Zhang W, Wang X R and Ni Z H 2015 Appl. Phys. Lett. 106 021121

[27] Gabor N M, Song J C W, Ma Q, Nair N L, Taychatanapat T, Watanabe K, Taniguchi T, Levitov L S and Jarillo-Herrero P 2011 Science 334 648

[28] Yan J, Kim M, Elle J A, Sushkov A B, Jenkins G S, Milchberg H M, Fuhrer M S and Drew H D 2012 Nat. Nanotechnol. 7 472

[29] Klekachev A V, Cantoro M, van der Veen M H, Stesmans A L, Heyns M M and Gendt S D 2011 Physica E 43 1046

[30] Guo W H, Xu S G, Wu Z F, Wang N, Loy M M T and Du S W 2013 Small 9 3031

[31] Qiao H, Yuan J, Xu Z Q, Chen C Y, Lin S H, Wang Y S, Song J C, Liu Y, Khan Q, Hoh H Y, Pan C X, Li S J and Bao Q L 2015 ACS Nano 9 1886

[32] Konstantatos G, Badioli M, Gaudreau L, Osmond J, Bernechea M, Arquer F P G

D, Gatti F and Koppens F H L 2012 Nat. Nanotechnol. 7 363

[33] Haider G, Roy P, Chiang C W, Tan W C, Liou Y R, Chang H T, Liang C T, Shih W H and Chen Y F 2016 Adv. Funct. Mater. 26 620

[34] Chen Z F, Cheng Z Z, Wang J Q, Wan X, Shu C, Tsang H K, Ho H P and Xu J B 2015 Adv. Opt. Mater. 3 1207

[35] Liu F Z and Kar S 2014 ACS Nano 8 10270

[36] Zeng L H, Wang M Z, Hu H, Nie B, Yu Y Q, Wu C Y, Wang L, Hu J G, Xie C, Liang F X and Luo L B 2013 ACS Appl. Mater. Interfaces 5 9362

[37] Guo X T, Wang W H, Nan H Y, Yu Y F, Jiang J, Zhao W W, Li J H, Zafar Z, Xiang N, Ni Z H, Hu W D, You Y M and Ni Z H 2016 Optica arXiv preprint arXiv:1608.05950

[38] Schroder D K 2001 Meas. Sci. Technol. 12 3

[39] Cuevas A, Macdonald D 2004 Sol. Energ. 76 255

[40] Liu X L, Luo X G, Nan H Y, Guo H, Wang P, Zhang L L, Zhou M M, Yang Z Y, Shi Y, Hu W D, Ni Z H, Qiu T, Yu Z F, Xu J B and Wang X R 2016 Adv. Mater. 28 5200

[41] Lee Y, Kwon J, Hwang E, Ra C H, Yoo W J, Ahn J H, Park J H and Cho J H 2015 Adv. Mater. 27 41

[42] Huisman, E H, Shulga A G, Zomer P J, Tombros N, Bartesaghi D, Bisri S Z, Loi M A, Koster L J A and van Wees B J 2015 ACS Appl. Mater. Interfaces 7 11083

[43] Tan W C, Shih W H and Chen Y F 2014 Adv. Funct. Mater. 24 6818

[44] Kim S J, Song W, Kim S, Kang M A, Myung S, Lee S S, Lim J and An K S 2016 Nanotechnology 27 075709

[45] Munoz E, Monroy E, Garrido J A, I. Izpura I, Sánchez F J, Sánchez-García M A, Calleja E, Beaumont B and Gibart P 1997 Appl. Phys. Lett. 71 870

[46] Xie F, Lu H, Xiu X Q, Chen D J, Han P, Zhang R and Zheng Y D 2011 Solid-State Electron. 57 39

[47] Perera A G U, Yuan H X, Choe J W, Francombe M H 1995 Proceedings of SPIE 2475 76

[48] Itkis M E, Niyogi S, Meng M E, Hamon M A, Hu H and Haddon R C 2002 Nano Lett. 2 155

[49] Lu R T, Christianson C, Weintrub B and Wu J Z 2013 ACS Appl. Mater. Interfaces 5 11703

[50] Liu Y D, Wang F Q, Wang X M, Wang X Z, Flahaut E, Liu X L, Li Y, Wang X R, Xu Y B, Shi Y and Zhang R 2015 Nat. Commun. 6 8589

[51] Xiu F X and Zhao T T 2013 Chin. Phys. B 22 96104
[52] Hasan M Z and Kane C L 2010 Rev. Mod. Phys 82 3045
[53] Liu C H, Chang Y C, Norris T B and Zhong Z H 2014 Nat. Nanotechnol. 9 273
[54] Mak K F, Lee C, Hone J, Shan J and Heinz T F 2010 Phys. Rev. Lett. 105 136805
[55] Radisavljevic B, Radenovic A, Brivio J, Giacometti V and Kis A 2011 Nat. Nanotechnol. 6 147
[56] Lopez-Sanchez O, Lembke D, Kayci M, Radenovic A and Kis A 2013 Nat. Nanotechnol. 8 497
[57] Wang H N, Zhang C J, Chan W M, Tiwari S and Rana F 2015 Nat. Commun. 6 8831
[58] Feng Y Q, Zhou W, Wang Y J, Zhou J, Liu E F, Fu Y J, Ni Z H, Wu X L, Yuan H T, Miao F, Wang B G, Wan X G and Xing D Y 2015 Phys. Rev. B 92 054110
[59] Tongay S, Sahin H, Ko C, Luce A, Fan W, Liu K, Zhou J, Huang Y S, Ho C H, Yan J Y, Ogletree D F, Aloni S, Ji J, Li S S, Li J B, Peeters F M and Wu J Q 2014 Nat. Commun. 5 3252
[60] Kim S, Konar A, Hwang W S, Lee J H, Lee J H, Yang J, Jung C, Kim H, Yoo J B, Choi J Y, Jin Y W, Lee S Y, Jena D, Choi W and Kim K 2012 Nat. Commun. 3 1011
[61] Liu E F, Long M S, Zeng J W, et al. 2016 Adv. Funct. Mater. 26 1938
[62] Wang X D, Wang P, Wang J L, Hu W D, Zhou X H, Guo N, Huang H, Sun S, Shen H, Lin T, Tang M H, Liao L, Jiang A Q, Sun J L, Meng X J, Chen X S, Lu W and Chu J H 2015 Adv. Mater. 27 6575
[63] Gong F, Luo W J, Wang J L, Wang P, Fang H H, Zheng D S, Guo N, Wang J L, Luo M, Ho J C, Chen X S, Lu W, Liao L and Hu W D 2016 Adv. Funct. Mater. DOI: 10.1002/adfm.201601346
[64] Xia F N, Wang H and Jia Y C 2014 Nat. Commun. 5 4458
[65] Huang M Q, Wang M L, Chen C, Ma Z W, Li X F, Han J B and Wu Y Q 2016 Adv. Mater. 28 3481
[66] Yuan H T, Liu X G, Afshinmanesh F, Li W, Xu G, Sun J, Lian B, Curto A G, Ye G J, Hikita Y, Shen Z X, Zhang S C, Chen X H, Brongersma M, Hwang H Y and Yi Cui 2015 Nat. Nanotechnol. 10 707
[67] Cho J H, Lee J, Xia Y, Kim B, He Y Y, Renn M J, Lodge T P and Frisbie C D 2008 Nat. Mater. 7 900
[68] Youngblood N, Chen C, Koester S J and Li M 2015 Nat. Photonics 9 247
[69] Zhang W J, Chuu, C P, Huang J K, Chen C H, Tsai M L, Chang Y H, Liang C T,

Chen Y Z, Chueh Y L, He J H, Chou M Y and Li L J 2014 Sci. Rep. 4 3826

[70] Yu W J, Liu Y, Zhou H L, Yin A X, Li Z, Huang Y and Duan X F 2013 Nat. Nanotechnol. 8 952

[71] Lee C H, Lee G H, Van Der Zande A M, Chen W C, Li Y L, Han M Y, Cui X, Arefe G, Nuckolls C, Heinz T F, Guo J, Hone J and Kim P 2014 Nat. Nanotechnol. 9 676

[72] Long M S, Liu E F, Wang P, et al. 2016 Nano Lett. 16 2254

[73] Tan C L and Zhang H 2015 Nat. Commun. 6 7873

[74] Hsieh T H, Lin H, Liu J W, Duan W H, Bansil A and Fu L 2012 Nat. Commun. 3 982

[75] Tanaka Y, Ren Z, Sato T, Nakayama K, Souma S, Takahashi T, Segawa K and Ando Y 2012 Nat. Phys. 8 800

[76] Xu S Y, Liu C, Alidoust N et al. 2012 Nat. Commun. 3 1192.

[77] Wang Q S, Xu K, Wang Z X, Wang F, Huang Y, Safdar M, Zhan X Y, Wang F M, Cheng Z Z and He J 2015 Nano Lett. 15 1183

[78] Wang Q S, Wen Y, Yao F R, Huang Y, Wang Z X, Li M L, Zhan X Y, Xu K, Wang F M, Wang F, Li J, Liu K H, Jiang C, Liu F Q and He J 2015 Small 11 5388

[79] Sun Z Q, Liao T, Dou Y H, Hwang S M, Park M S, Jiang L, Kim J H and Dou S X 2014 Nat. Commun. 5 3813

[80] Mak K F, He K L, Lee C, Lee G H, Hone J, Heinz T F and Shan J 2013 Nature Mater. 12 207

[81] Fogler M M, Butov L V and Novoselov K S 2014 Nature Commun. 5 4555

[82] Shi H Y, Yan R S, Bertolazzi S, Brivio J, Gao B, Kis A, Jena D, Xing H G and Huang L B 2013 ACS Nano 7 1072

[83] Mak K F and Shan J 2016 Nat. Photonics 10 216

[84] Tongay S, Suh J, Ataca C, Fan W, Luce A, Kang J S, Liu J, Ko C, Raghunathanan R, Zhou J, Ogletree F, Li J B, Grossman J C and Wu J Q 2013 Sci. Rep. 3 2657

[85] Baugher B W H, Churchill H O H, Yang Y F and Jarillo-Herrero P 2014 Nat. Nanotechnol. 9 262

[86] Pospischil A, Furchi M M and Mueller T 2014 Nat. Nanotechnol. 9 257

[87] Ross J S, Klement P, Jones A M, Ghimire N J, Yan J Q, Mandrus D G, Taniguchi T, Watanabe K, Kitamura K, Yao W, Cobden D H and Xu X D 2014 Nat. Nanotechnol. 9 268

[88] Jo S, Ubrig N, Berger H, Kuzmenko A B and Morpurgo A F 2014 Nano Lett. 14

2019

[89] Zhang Y J, Oka T, Suzuki R, Ye J T and Iwasa Y 2014 Science 344 725

[90] Ye Y, Ye Z L, Gharghi M, Zhu H Y, Zhao M, Wang Y, Yin X B and Zhang X 2014 Appl. Phys. Lett. 104 193508

[91] Yang W H, Shang J Z, Wang J P, Shen X N, Cao B C, Peimyoo N, Zou C J, Chen Y, Wang Y L, Cong C X, Huang W and Yu T 2016 Nano Lett. 16 1560

[92] Cheng R, Li D H, Zhou H L, Wang C, Yin A X, Jiang S, Liu Y, Chen Y, Huang Y and Duan X F 2014 Nano Lett. 14 5590

[93] Withers F, Pozo-Zamudio O D, Mishchenko A, Rooney A P, Gholinia A, Watanabe K, Taniguchi T, Haigh S J, Geim A K, Tartakovskii A I and Novoselov K S 2015 Nat. Mater. 14 301

[94] Politi A, Matthews J C F and O'Brien J L 2009 Science 325 1221

[95] He Y M, Clark G, Schaibley J R, He Y, Chen M C, Wei Y J, Ding X, Zhang Q, Yao W, Xu X D, Lu C Y and Pan J W 2015 Nat. Nanotechnol. 10 497

[96] Koperski, Nogajewski M K, Arora A, Cherkez V, Mallet P, Veuillen J Y, Marcus J, Kossacki P and Potemski M 2015 Nat. Nanotechnol. 10 503

[97] Chakraborty C, Kinnischtzke L, Goodfellow K M, Beams R and Vamivakas A N 2015 Nat. Nanotechnol. 10 507

[98] Srivastava A, Sidler M, Allain A V, Lembke D S, Kis A and Imamoğlu A 2015 Nat. Nanotechnol. 10 491

[99] Tran T T, Bray K, Ford M J, Toth M and Aharonovich I 2016 Nat. Nanotechnol. 11 37

# Figures

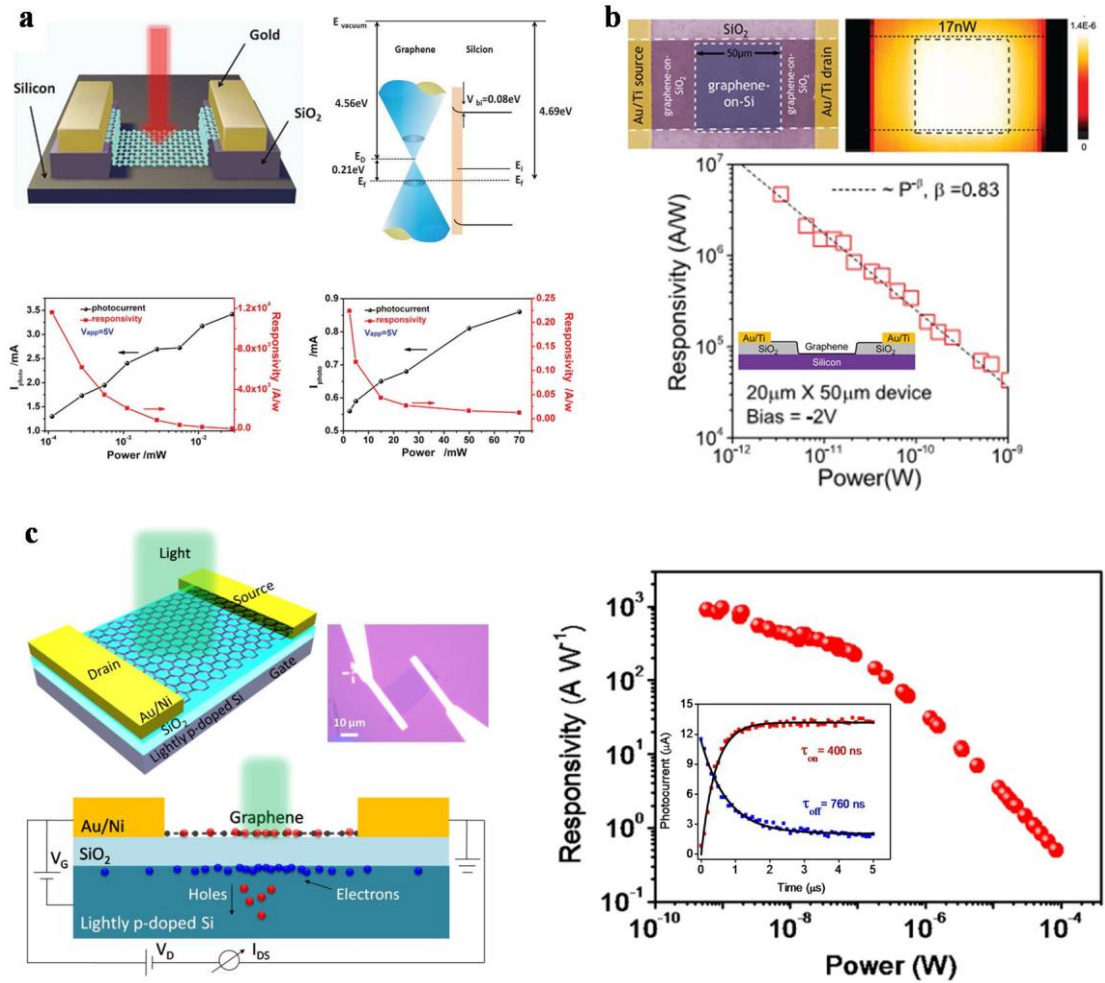

**Fig. 1.** (a) Schematic of the silicon/graphene photodetector (top left). The energy band diagram of silicon/graphene heterostructure (top right). The graphs at bottom display photocurrent and photoresponsivity versus illumination power at wavelength of 632 nm (bottom left) and 1550 nm (bottom right), respectively. (from Ref. [34]) (b) A profile schematic and a pseudo-colored SEM image of the silicon/graphene high-gain photodetector (top left). Photocurrent map at incident power of 17 nW shows bright photocurrent signal in the silicon/graphene region (top right). Photoresponsivity of the device (inset) as a function of incident power (bottom). (from Ref. [35]) (c) Schematic diagram and optical image of the graphene hybrid photodetector on lightly p-doped silicon/$SiO_2$ substrate(left). Responsivity at $V_D = 1$ V and $V_G = 0$ V of the device as a function of the light power(right). The transient response of the device switched on or off by an acoustic optical modulator with frequency of 10 kHz. P= ~0.5 μW, $V_D$= 1 V and $V_G$= 0 V (inset). (from Ref. [37])

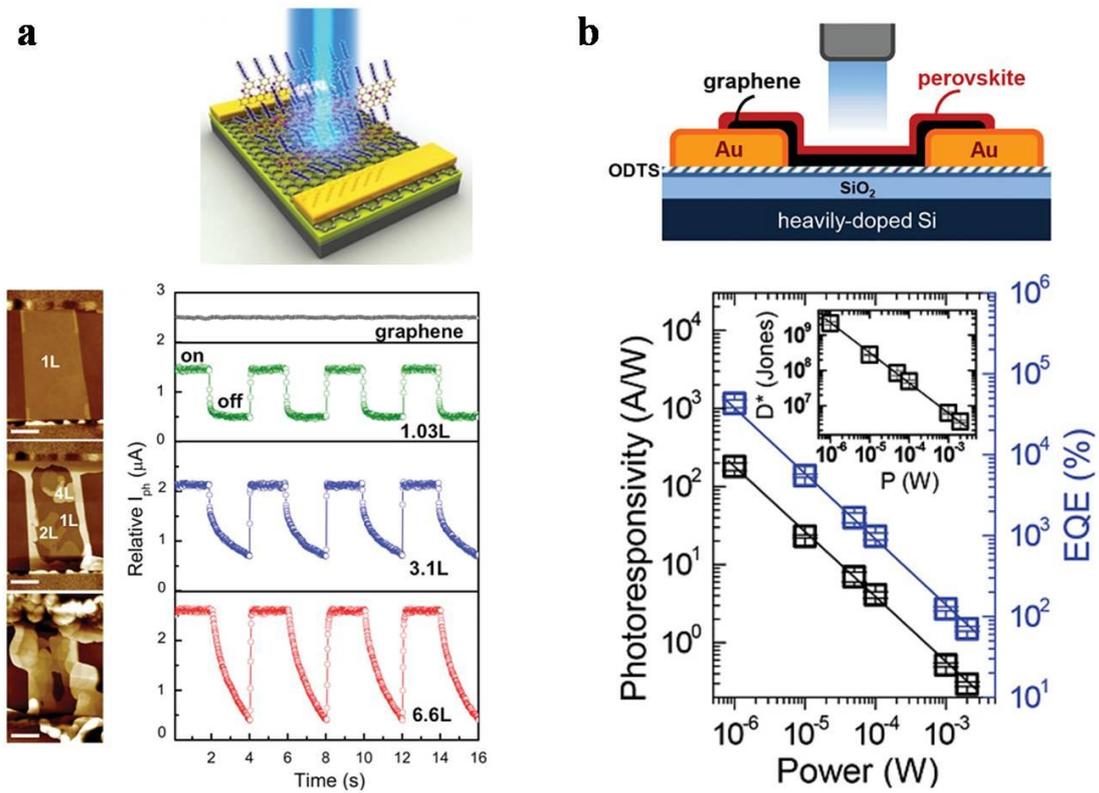

**Fig. 2.** (a) Schematic of a typical organic materials/graphene hybrid phototransistor (top). Atomic force microscope (AFM) images of a device undergone repeated $C_8$-BTBT growth (bottom left). Photocurrent response of the graphene and $C_8$-BTBT/graphene hybrid devices under the same experimental conditions: laser power density 7000 μWcm$^{-2}$, bias voltage 0.1V, $V_g$-$V_0$=10 V. (bottom right). (from Ref. [40]) (b) Schematic of the perovskite/graphene hybrid photodetector (top). Photoresponsivity and EQE vs. illumination power of the perovskite/graphene hybrid photodetector (bottom). The photodetectivity (D*) vs. illumination power (inset). (from Ref. [41])

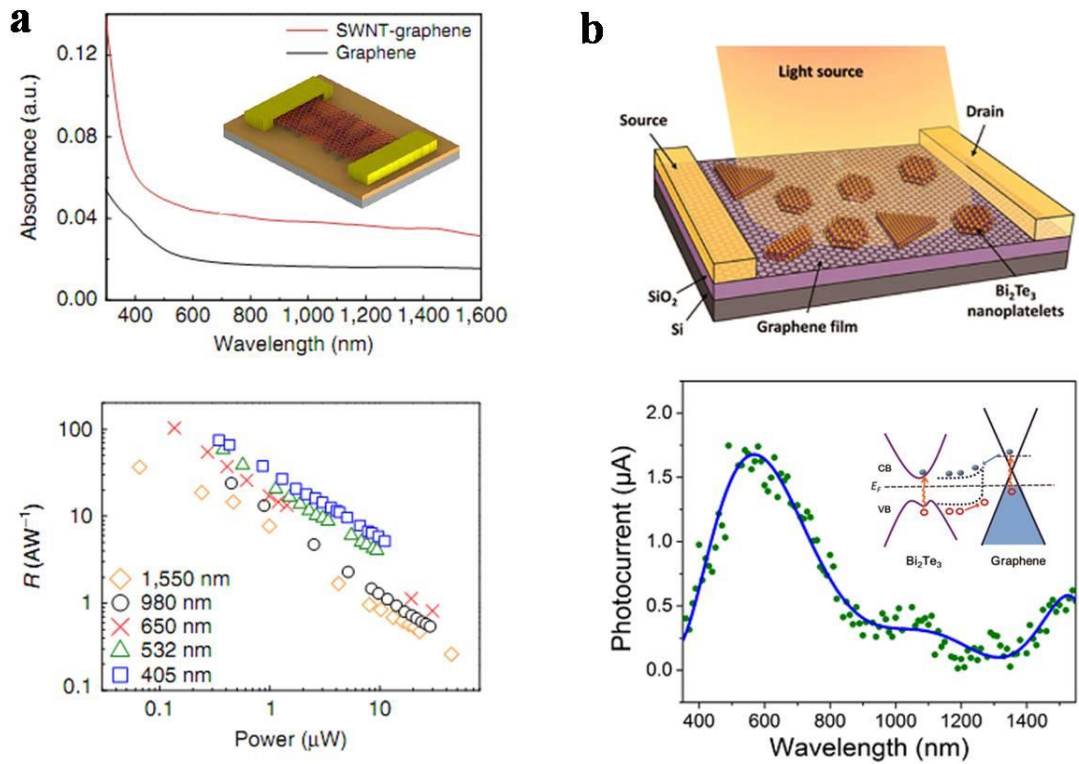

**Fig. 3.** (a) The ultraviolet-visible-infrared light absorbance of graphene and SWCNT/graphene hybrid film on quartz (top). Responsivities of SWCNT/graphene photodetector as a function of the optical power at different illumination wavelengths (bottom). (from Ref. [50]) (b) Schematic of the $Bi_2Te_3$/graphene hybrid device (top). Photocurrent as a function of wavelength (bottom). Illustration of band structure and charge transfer (inset). (from Ref. [31])

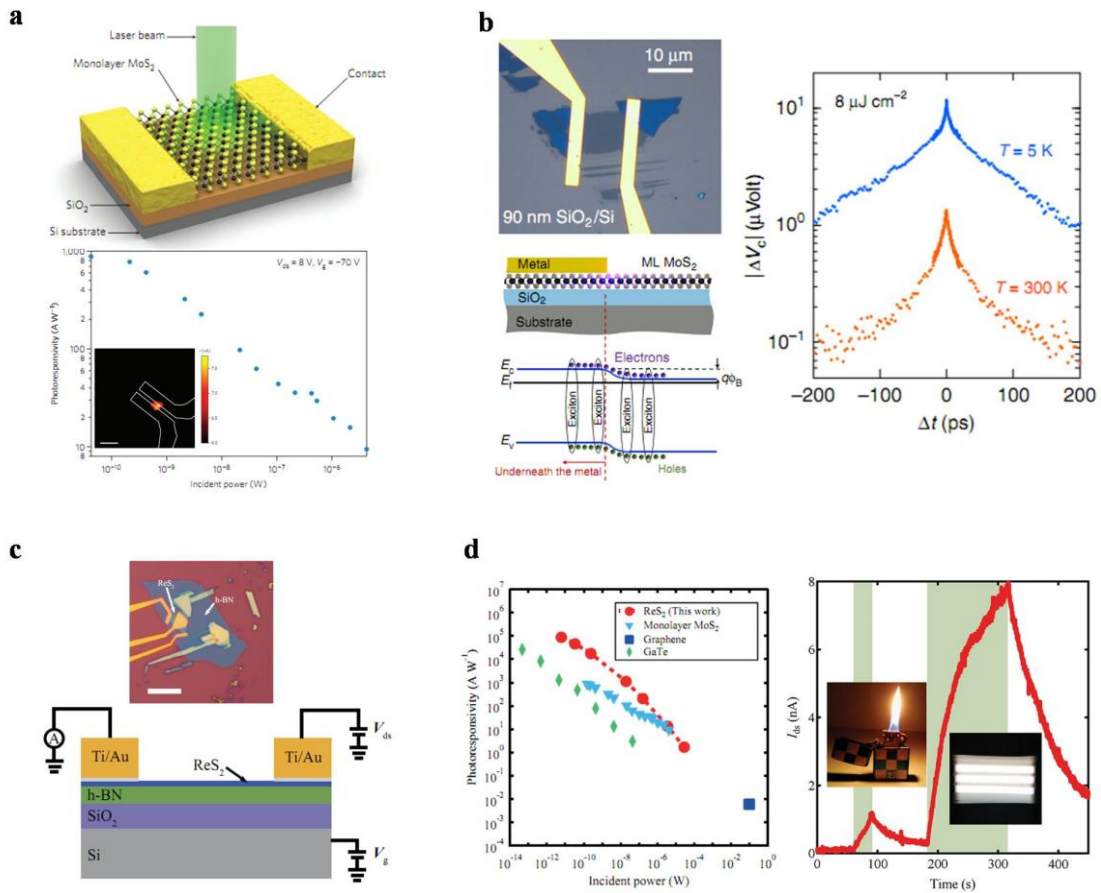

**Fig. 4.** (a) Cross-sectional view of the single-layer MoS$_2$ based photodetector under the focused laser beam (top). Photoresponsivity of the MoS$_2$ phototransistor (bottom). Spatial map of the photocurrent of the device (inset). (from Ref. [56]) (b) Optical image of a back-gated monolayer MoS$_2$ photodetector on SiO$_2$/silicon substrate (top left). The energy band diagram (bottom) of the metal-MoS$_2$ junction (top) after photoexcitation with an optical pulse (bottom left). The photoresponse time characterization (right). The plots show two distinct timescales: a fast timescale of ~4.3 ps and a slow timescale of ~105 ps. (from Ref. [57]) (c) Schematic and measurement circuit of the few-layer ReS$_2$ phototransistors. (d) Photoresponsivity as a function of the incident power of few-layer ReS$_2$ phototransistors (left). Weak signal detection in a five-layer ReS$_2$ phototransistor using lighter and limited fluorescent lighting as the weak light sources (right). ((c) and (d) are reproduced from Ref. [61])

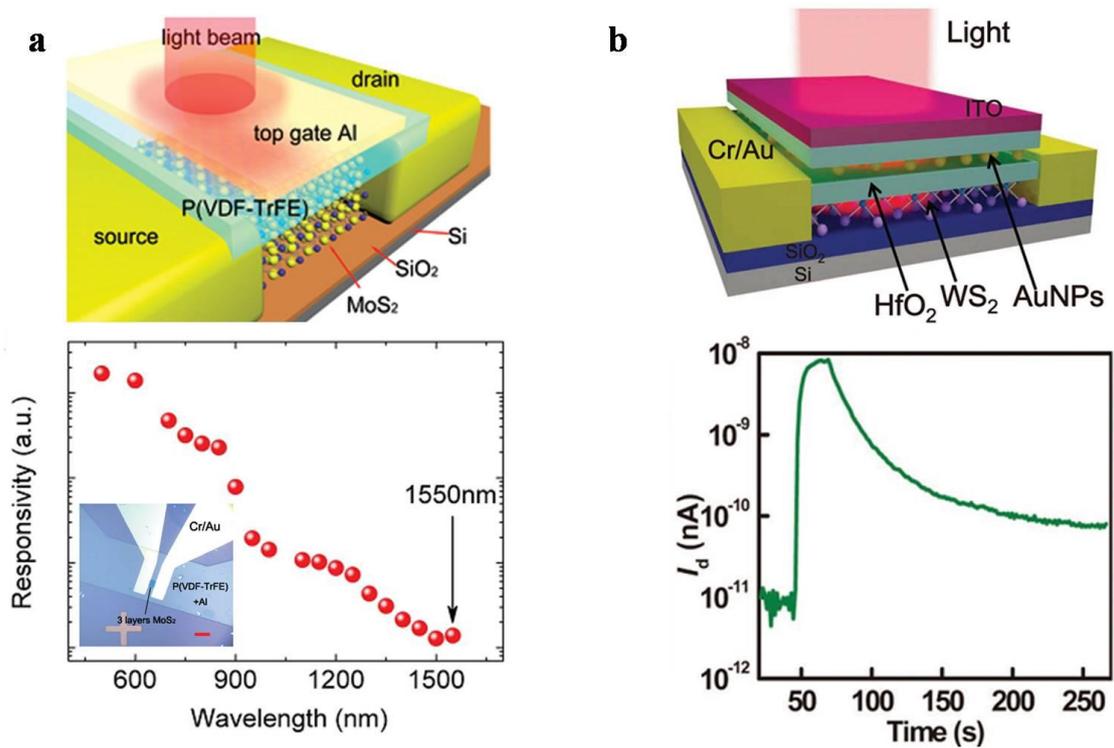

**Fig. 5.** (a) Schematic of the triple-layer MoS$_2$ photodetector with monochromatic light beam (top). The responsivity of the device at different wavelength (bottom). Optical image of the device (inset). (from Ref. [62]) (b) Schematic of the WS$_2$ phototransistor with gold nanoparticles (AuNPs) embedded in the gate dielectric (top). Normalized current response at continuous on/off cycles of light illumination (bottom). (from Ref. [63])

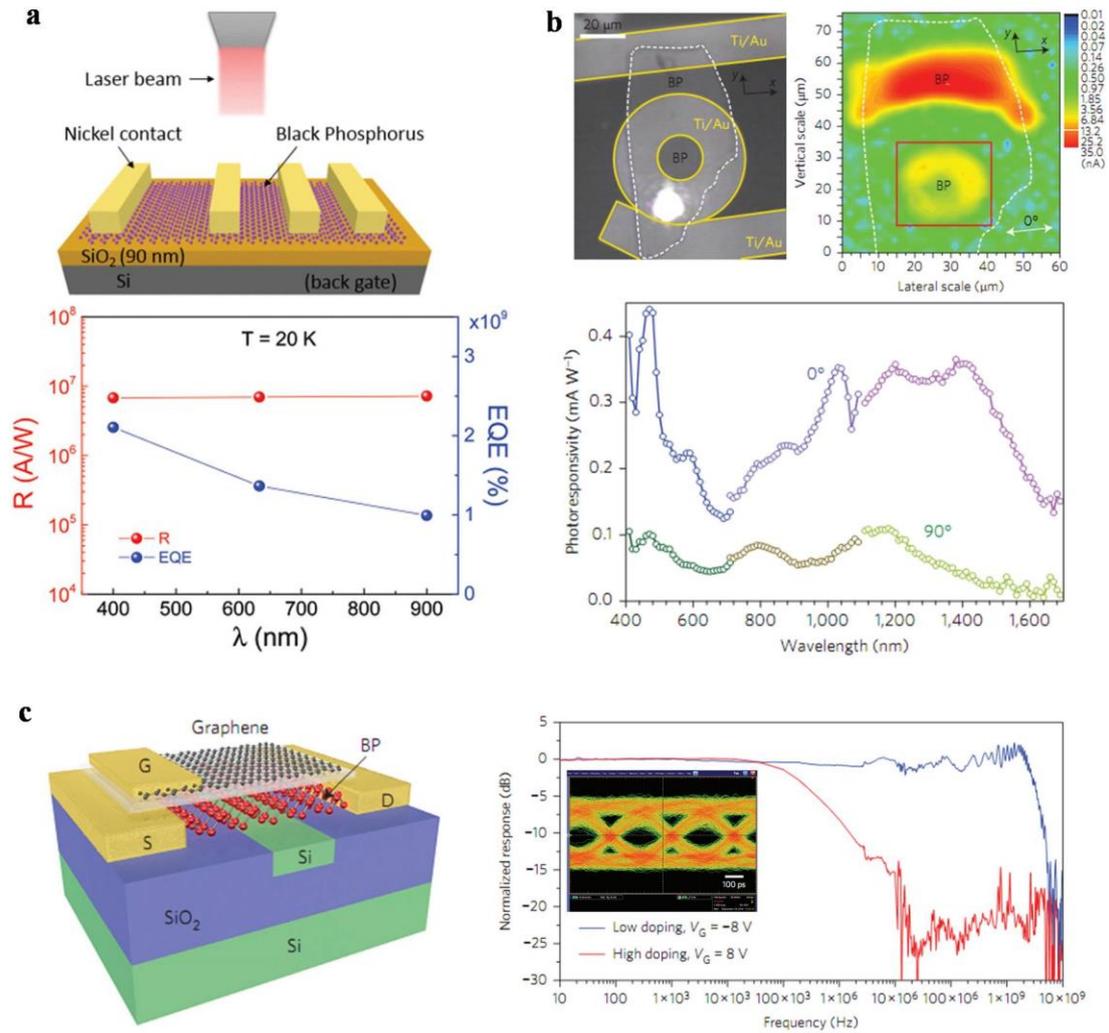

**Fig. 6.** (a) Schematic of a BP photodetector (top). Photoresponsivity and EQE measured at different wavelengths (bottom). (from Ref. [65]) (b) Optical image of a BP photodetector. Instead of a straight-edge metal electrode, the isotropic round photocurrent collector was employed to avoid the linear polarization that might arise from a straight metal edge (top left). Corresponding photocurrent microscopy image of the device, with illumination at 1500 nm and polarization along the direction of the white arrow (x axis) (top right). Polarization dependence of photoresponsivity with illumination from 400 to 1,700 nm, where the polarization angle of 0° corresponds to the x crystal axis and 90° corresponds to the y crystal axis (bottom). (from Ref. [66]) (c) Schematic of the waveguide integrated BP photodetector (left). The response of the BP photodetector is measured when BP is gated to low and high doping conditions (right). Receiver eye diagram at a data rate of 3 GHz (inset). (from Ref. [68])

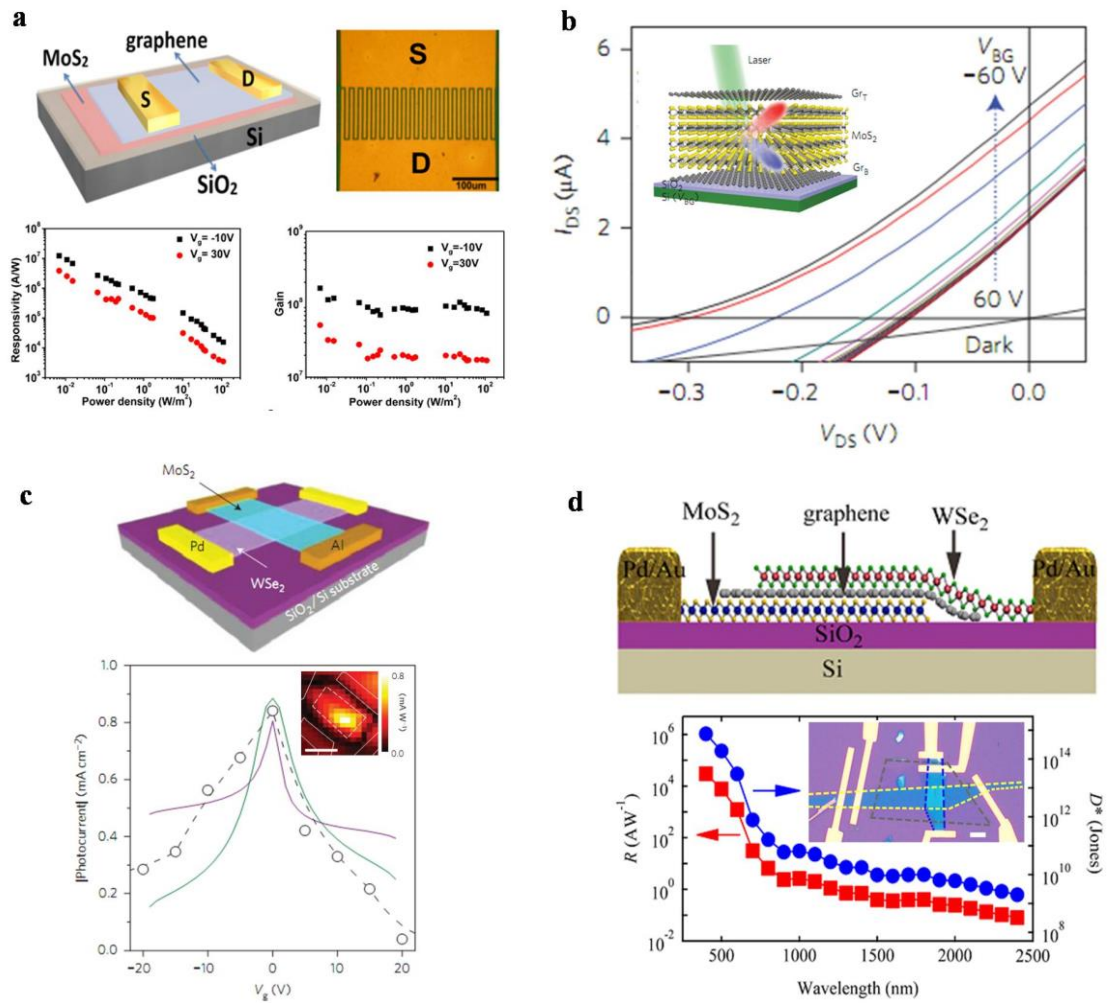

**Fig. 7.** (a) Schematic illustration (top left) of the photodetector based on graphene/MoS$_2$ heterostructure, where the channel is formed in between the comb-shaped source and drain metal electrodes (top right). Photoresponsivity (bottom left) and photogain (bottom right) for the graphene/MoS$_2$ photodetectors. (from Ref. [69]) (b) I-V characteristics of the graphene-MoS$_2$-graphene heterostructures based device under laser illumination. Schematic illustration of the side view of the device (inset). (from Ref. [70]) (c) Schematic diagram of a van der Waals-stacked MoS$_2$/WSe$_2$ heterojunction device with lateral metal contacts (top). Measured and simulated photocurrent at V$_{ds}$ = 0 V as a function of gate voltages (bottom). Photocurrent map of the device (inset). (from Ref. [71]) (d) Cross-section schematic of MoS$_2$-graphene-WSe$_2$ heterostructure based photodetector (top). Photoresponsivity R (left) and specific detectivity D* (right) of a typical device for wavelengths range from 400 to 2400 nm measured in ambient air (bottom). Optical image of the device (inset). (from Ref. [72])

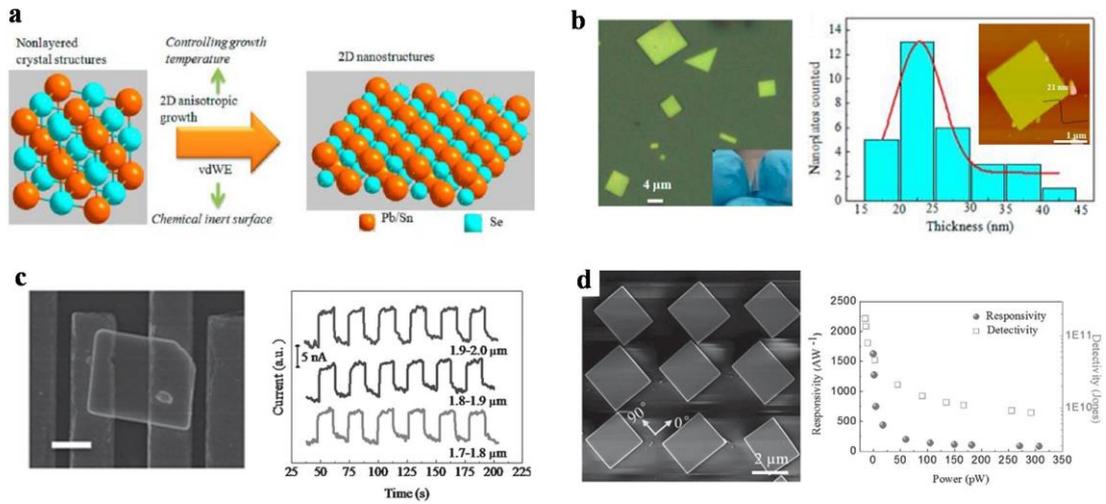

**Fig. 8.** (a) $Pb_{1-x}Sn_xSe$ cubic crystal structure can be tailored into ultrathin 2D nanostructure by 2D anisotropic growth. (b) Optical image of $Pb_{1-x}Sn_xSe$ nanoplates (left). Histogram of $Pb_{1-x}Sn_xSe$ nanoplates thickness (right). AFM image of the nanoplate (inset). ((a) and (b) are reproduced from Ref. [77]) (c) SEM image of the $Pb_{1-x}Sn_xSe$ nanoplates based device (left). Mid-infrared photodetection of $Pb_{1-x}Sn_xSe$ nanoplates at 1.7-2.0 μm (right). (from Ref. [78]) (d) SEM image of PbS nanoplate arrays (left). The photoresponsivity and detectivity of the PbS nanoplates device as a function of laser power. The wavelength of incident laser was 800 nm (right). (from Ref. [13])

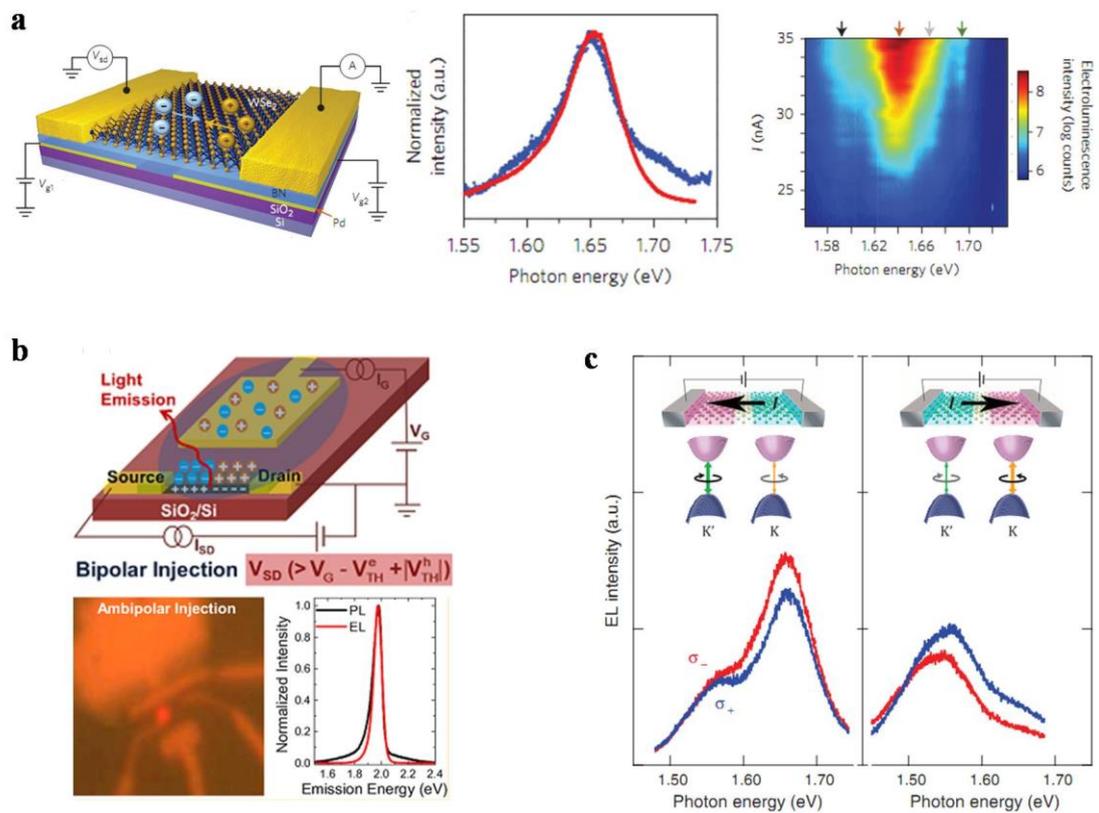

**Fig. 9.** (a) Schematic of monolayer WSe$_2$ p-n junction devices with palladium backgates ($V_{g1}$ and $V_{g2}$) and source (S) and drain (D) contacts (left). Photoluminescence and electroluminescence of the LEDs (middle). Electroluminescence intensity as a function of bias current and photon energy (right). From left to right, the arrows indicate the impurity-bound exciton, the negative and positive charged excitons and the neutral exciton. (from Ref. [87]) (b) Schematic illustrations of an ionic liquid gated device in the unipolar injection regime (top). Optical microscope images of the monolayer device with clearly visible spot of emitted light due to electroluminescence (bottom left). Photoluminescence and electroluminescence spectra of the LEDs (bottom right). (from Ref. [88]) (c) Circularly polarized electroluminescence spectra from a polarized WSe$_2$-based LED for two opposite current directions (top). The contribution to electroluminescence from two valleys (bottom). (from Ref. [89])

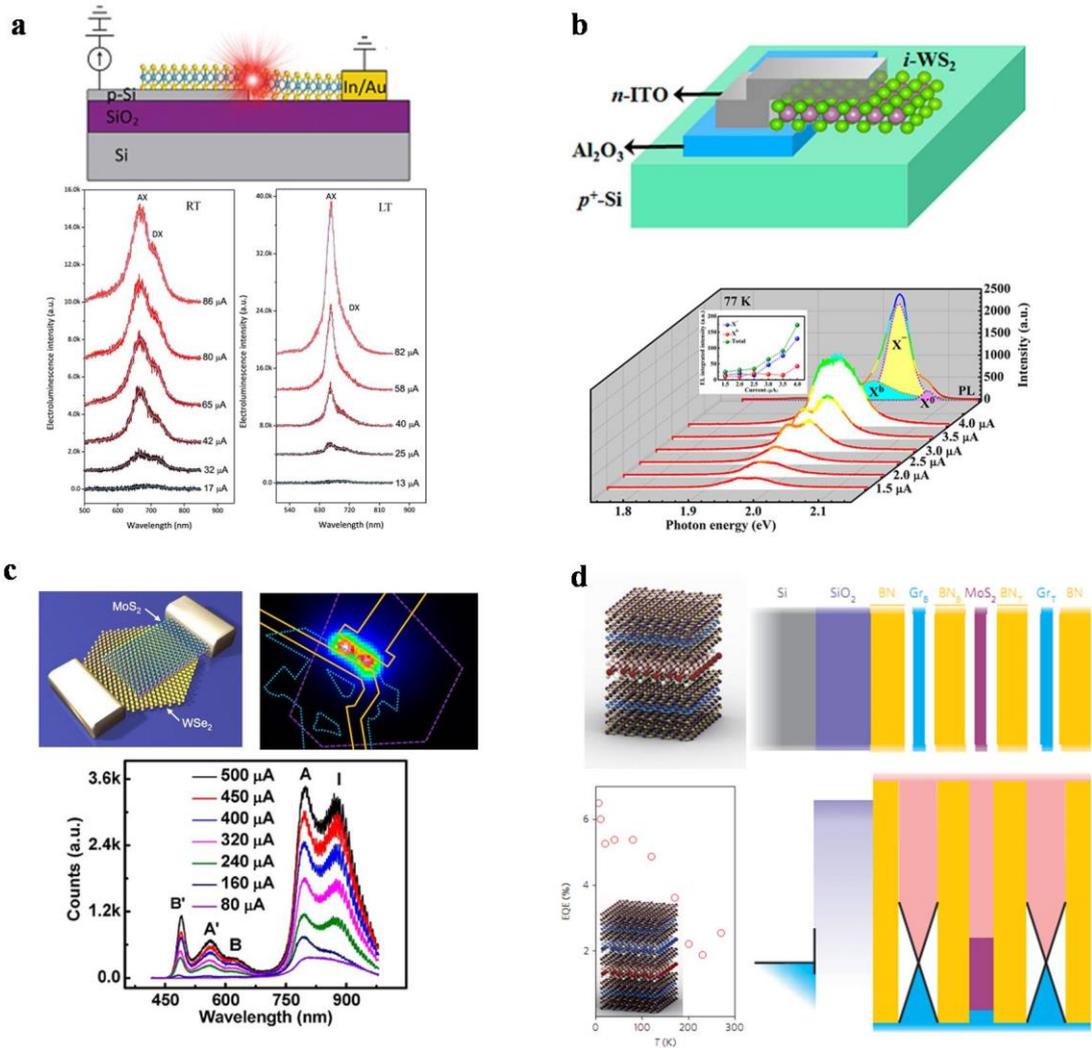

**Fig. 10.** (a) Schematic of the MoS$_2$/silicon heterojunction electroluminescence device (top). Electroluminescence spectra of a MoS$_2$ diode at room temperature (bottom left) and low temperature (bottom right), respectively. (from Ref. [90]) (b) Schematic of the p+-Si/i-WS$_2$/n-ITO heterojunction LED device (top). Electroluminescence spectra of the LED device recorded at 77 K under different injection current (bottom). (from Ref. [91]) (c) Schematic illustration (top left) and electroluminescence image (top right) of the WSe$_2$/MoS$_2$ heterojunction p−n diode. The electroluminescence spectra of a bilayer WSe$_2$/MoS$_2$ heterojunction at different injection current (bottom). (from Ref. [92]) (d) Schematic of the single-quantum-well (SQW) heterostructure (top left). Schematic of the heterostructure consisting of Si/SiO$_2$/hBN/GrB/3hBN/MoS$_2$/3hBN/GrT/hBN (top right). Temperature dependence of EQE for a device with two quantum wells (QWs) made from MoS$_2$ and WSe$_2$ (bottom left). Band diagrams for the case of zero applied bias for the heterostructure showed in the top right panel (bottom right). (from Ref. [93])

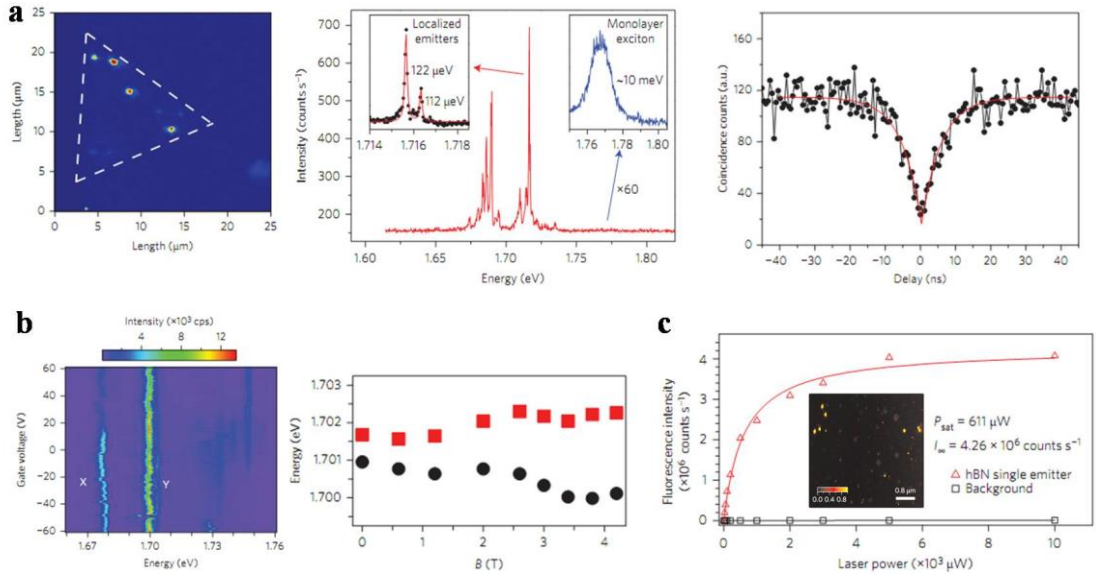

**Fig. 11.** (a) Photoluminescence image of isolated defect QDs in WSe$_2$. The dashed triangle indicates the position of the monolayer WSe$_2$ (left). Photoluminescence spectrum of localized emitters (middle). The emission of the localized emitters exhibits a much sharper spectral lines (the linewiths of the two spectra are ~112 and 122 μeV, respectively. left inset) than that of free excitons (~10 meV, right inset). Second-order correlation measurements revealed a strong photon antibunching (right). (from Ref. [95]) (b) WSe$_2$ emission spectra as a function of applied gate voltage, corresponding to light emission from quantum dot-like defects (left). Magnetic field dependence of a quantum dot emission spectra in the Faraday configuration (right). (from Ref. [97]) (c) Fluorescence saturation curve obtained from a single defect in hBN multilyers. Scanning confocal image of a multilayer hBN sample (inset). (from Ref. [99])